\documentclass[twocolumn,printnumbers,amsmath,amssymb]{revtex4}
\usepackage{graphicx}
\usepackage{color}

\begin{document}

\title{Low-frequency vibrational modes of stable glasses}

\date{\today}

\author{Lijin Wang$^{1,2}$}
\author{Andrea Ninarello$^{3}$}
\author{Pengfei Guan$^{1}$}
\author{Ludovic Berthier$^{3}$}
\author{Grzegorz Szamel$^{2}$}
\author{Elijah Flenner$^{2}$}

\affiliation{$^1$Beijing Computational Science Research Center, Beijing 100193, P. R. China}

\affiliation{$^2$Department of Chemistry, Colorado State University, Fort Collins, Colorado 80523, USA}

\affiliation{$^3$Laboratoire Charles Coulomb (L2C), University of Montpellier, CNRS, 34095 Montpellier, France}

\begin{abstract}
\textbf{
Unusual features of the vibrational density of states $D(\omega)$ of glasses allow one to rationalize their peculiar
low-temperature properties. Simulational studies of $D(\omega)$ have
been restricted to studying poorly annealed glasses that may not be relevant to experiments.
Here we report on $D(\omega)$ of zero-temperature glasses with kinetic stabilities ranging from poorly annealed
to ultrastable glasses.
For all preparations, the low-frequency part of $D(\omega)$ splits between extended and quasi-localized modes. Extended modes exhibit a boson peak crossing over to Debye behavior ($D_\text{ex}(\omega) \sim \omega^2$) at low-frequency, with a strong correlation between the two regimes. Quasi-localized modes obey $D_\text{loc}(\omega) \sim \omega^4$, irrespective of the stability. The prefactor of this quartic law decreases with increasing stability, and the corresponding modes become more localized and sparser. Our work is the first numerical observation of quasi-localized modes in a regime relevant to experiments, and it establishes a direct connection between glasses' stability and their soft vibrational modes.}
\end{abstract}

\maketitle

Amorphous solids exhibit universal low-temperature properties, seen for instance in the heat capacity and thermal conductivity \cite{Zeller1971}, that differ remarkably from crystal physics. These properties are related to the vibrational density of states $D(\omega)$. For a continuous elastic medium in three dimensions, low-frequency excitations are phonons, and the density of states follows $D(\omega) = A_{\mathrm{D}} \omega^2$ where $A_{\mathrm{D}}$ is given by Debye theory~\cite{kittel}. A well-known universal feature of amorphous solids is an excess in vibrational modes over the Debye prediction that results in a peak in $D(\omega)/\omega^2$ at an intermediate frequency, called the boson peak~\cite{nakayama,parisi_nature,Inoue1991,Buchenau2007}.

More recently, another source of `excess modes' has been identified in computer simulations of model glasses~\cite{lerner_prl2016,ikeda_pnas,lerner_pre2017,ikeda_pre2018,lerner2018,Angelani2018}.
It is composed of quasi-localized low-frequency modes with a density obeying $D_{\mathrm{loc}}(\omega) \sim \omega^4$. Quasi-localized modes are observed at frequencies significantly lower than the boson peak and the link between the two phenomena is not immediate, despite some indications that they may be connected~\cite{lerner2018b,ikeda_pnas}.
The quartic law was predicted long ago using phenomenological models~\cite{Buchenau1991,Buchenau1992}, reanalyzed over the years~\cite{Schober1996,Gurevich_prb2003,Schrimacher_prl2007}, and remains the focus of intense research~\cite{fernanda,lisa}. These predictions differ from two recent mean-field approaches~\cite{meanfield1,meanfield2}, which predict instead a universal non-Debye behavior that is quadratic in all spatial dimensions, also reported numerically~\cite{patrick}.
Interest in the low-frequency localized modes extends beyond connections to theoretical models and the boson peak. It was suggested that these modes are correlated with irreversible structural relaxation in the supercooled liquid state~\cite{Widmer-Cooper_np}, and that the spatial distribution of these soft modes is correlated with rearrangements upon mechanical deformation and plasticity~\cite{rottler_prx,chen_prl,manning_prl,Zylberg_PNAS2017}. Localized defects are also central to theoretical descriptions of glass properties at cryogenic temperatures~\cite{somereftls1,somereftls2}.

Recent numerical insights were obtained for glasses that are very different from the ones studied experimentally, since they are prepared with protocols operating on timescales that differ from experimental ones by as many as ten orders of magnitude~\cite{berthier_rmp}. It is therefore unknown whether any of the vibrational, thermal, or mechanical properties derived from earlier computational study of the density of states is experimentally relevant.
For example, it was reported \cite{lerner_pre2017,Lerner2018} that
$D_{\mathrm{loc}}(\omega) \sim \omega^\beta$ with $\beta$ ranging from 3 to 4 depending on the
glass's stability, with $\beta = 4$ for the two most stable simulated glasses
created by cooling at a constant rate. It remains unclear, however, whether $\beta$ would be different for
glasses with stability comparable to that of the experimental glasses. 

Our main achievement  is to extend studies of the vibrational density of states of computer glasses to an experimentally-relevant regime of glass stability for the first time. To this end, we build on the recent development of a Monte Carlo method that allows us to equilibrate supercooled liquids down to temperatures below the experimental glass transition~\cite{berthier_prx2017,berthier_prl2016,chris2017} to prepare glasses that cover an unprecedented range of kinetic stability, from extremely poorly-annealed systems to ultrastable glasses. We thus match the large gap between previous numerical findings and the experimental regime~\cite{berthier_pnas2017}. Recent studies have shown that that such stable glasses may differ qualitatively from ordinary computer glasses~\cite{chris2017,scalliet2017,misaki}. For example, qualitatively different yielding behavior of well annealed
glasses compared to that of poorly annealed glasses was reported in Ref. \cite{misaki}.
Since rearrangements upon mechanical deformation are correlated with the spatial distribution of soft modes,
this result suggested that the density of states could also evolve dramatically with the stability. 

\section*{Results}
\textbf{System preparation.}
We prepare glasses by instantaneously quenching supercooled liquids equilibrated at parent temperature $T_{\mathrm{p}}$ to $T=0$, so that $T_{\mathrm{p}}$ uniquely controls the glass stability. We find that the low-frequency part of the vibrational density of states changes considerably when $T_{\mathrm{p}}$ varies, thus offering a direct link between soft vibrational modes and kinetic stability. Following earlier work~\cite{ikeda_pnas,ikeda_pre2018}, we divide modes into extended and quasi-localized ones. As found for high parent temperature glasses~\cite{lerner_prl2016,ikeda_pnas,lerner_pre2017,ikeda_pre2018}, the density of states of the quasi-localized modes follows  $D_{\mathrm{loc}} = A_4 \omega^4$, with the same quartic exponent for all glass stabilities.
Our work thus establishes the relevance of earlier findings about quasi-localised modes and their effect on the density of states in the experimentally relevant regime of glass stability. In addition, we find that the overall scale $A_4$ decreases surprisingly rapidly when $T_{\mathrm{p}}$ decreases, showing that the density of the quasi-localized modes is highly sensitive to the glass stability. This rapid decrease contrasts with the modest changes found for other structural quantities, such as mechanical moduli, sound speed, and Debye frequency. Quasi-localized modes also become sparser and increasingly localized at low $T_{\mathrm{p}}$, and so the identification of soft localized modes as relevant glassy defects controlling the physics of amorphous solids becomes more convincing near the experimental glass transition. Our results also suggest that ultrastable glasses contain significantly fewer localized excitations than ordinary glasses, which appears consistent with recent experiments~\cite{specific_heat_pnas,hellman,Ramos2015}.

\begin{figure}
\includegraphics[width=0.48\textwidth]{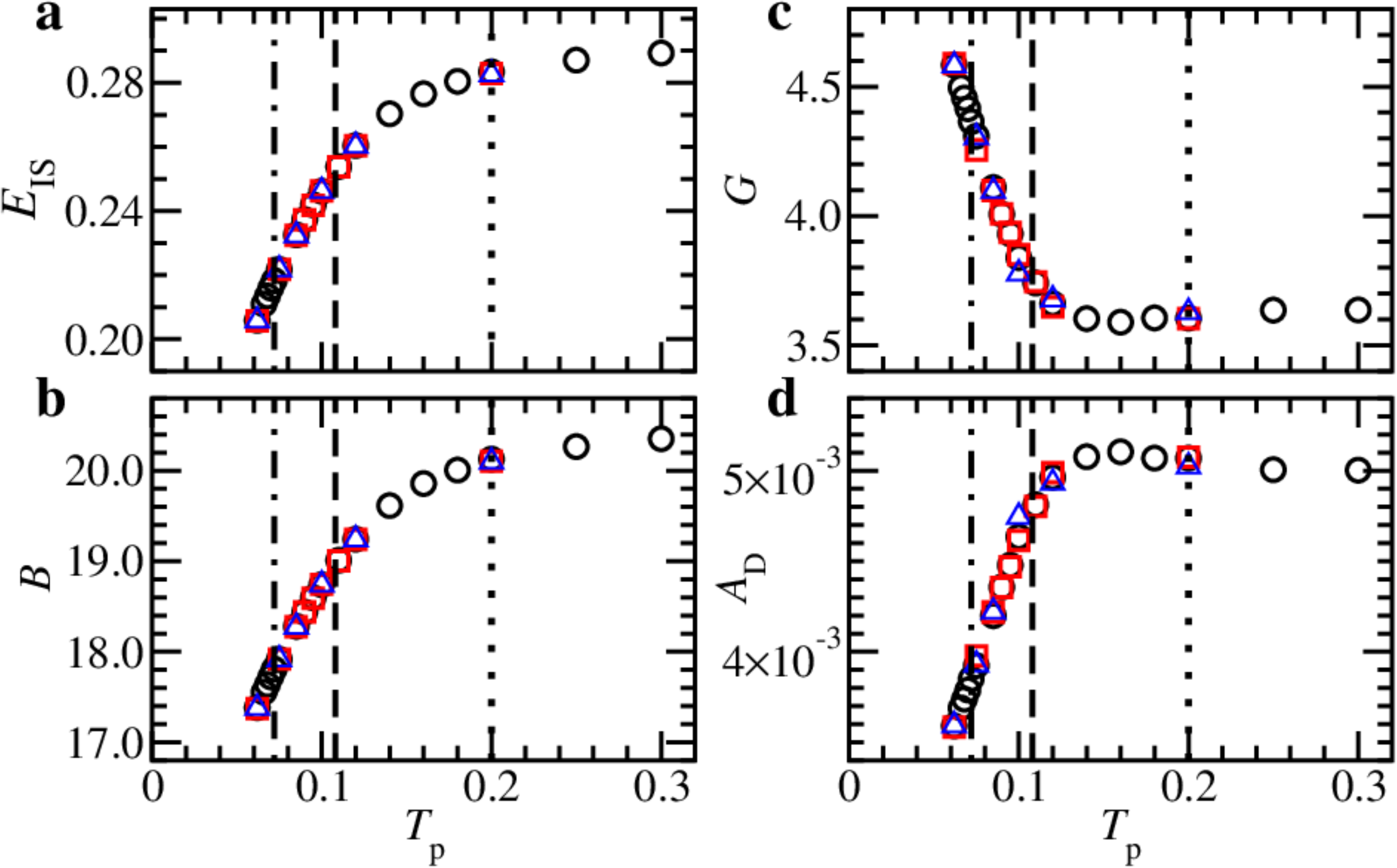}
\caption{\label{macro} \textbf{Macroscopic properties}.
\textbf{a} Inherent structure energy $E_{\mathrm{IS}}$ ;
\textbf{b} Bulk modulus $B$;
\textbf{c} Shear modulus $G$;
\textbf{d} Debye level $A_{\mathrm{D}}$.
In all panels, the vertical dashed-dotted, dashed and dotted lines mark the positions of $T_{\mathrm{g}}$, $T_{\mathrm{c}}$
and $T_{\mathrm{o}}$, respectively. Data shown are for $N=48000$ (circles), $96000$ (squares), and 192000 (triangles).}
\end{figure}

We simulate a polydisperse glass forming system in three dimensions, which is a representative glass-forming computer model~\cite{berthier_prx2017}.
We use the swap Monte Carlo algorithm to prepare independent equilibrated configurations at parent temperatures $T_{\mathrm{p}}$ ranging from
above the onset temperature of slow dynamics $T_{\mathrm{o}} \approx 0.200$, down to $T_{\mathrm{p}} = 0.062$, which is about $60\%$ of the mode-coupling temperature
$T_{\mathrm{c}} \approx 0.108$ ($T_{\mathrm{c}}$ marks  a crossover to activated dynamics and corresponds typically to the lowest temperature accessed by standard molecular dynamics). Importantly, our lowest $T_{\mathrm{p}}$ is lower than the estimated experimental glass temperature $T_{\mathrm{g}} \approx 0.072$~\cite{berthier_prx2017}, and no previous computational study has explored such range of glass stability.
In addition, we also use a very high parent temperature which we refer to as $T_{\mathrm{p}}=\infty$. We then probe vibrational properties of zero-temperature glasses
produced by an instantaneous quench from equilibrated configurations at different $T_{\mathrm{p}}$. The specific simulation details are provided in Methods.

\textbf{ Macroscopic  properties.} We begin
by presenting macroscopic properties of the glasses as a function of the parent temperature $T_{\mathrm{p}}$.
The inherent structure energy $E_{\mathrm{IS}}$ is directly related to the mobility of the particles \cite{Helfferich2016}, and thus we
show $E_{\mathrm{IS}}$ in Fig.~\ref{macro}a as an indicator of the increased stability of the glass.
$E_{\mathrm{IS}}$ deviates from its high temperature plateau when $T_{\mathrm{p}}$ becomes smaller than the onset temperature,
and decreases further with decreasing $T_{\mathrm{p}}$~\cite{sastry_nature_1998}.
Similarly, the bulk modulus $B$ decreases modestly with decreasing $T_{\mathrm{p}}$,
Fig.~\ref{macro}b. By contrast, the shear modulus $G$ in Fig.~\ref{macro}c remains nearly temperature-independent until the mode-coupling temperature,
which is below the onset temperature,  and then the shear modulus increases
with decreasing $T_{\mathrm{p}}$.
Associated with the increase in the shear modulus is a decrease in the Debye level $A_{\mathrm{D}} = 3/\omega_{\mathrm{D}}^3$ where the
Debye frequency $\omega_{\mathrm{D}}  = [(18 \pi^2 \rho)/(c_{\mathrm{l}}^{-3} + 2 c_{\mathrm{t}}^{-3}]^{1/3}$.
The decrease of $A_{\mathrm{D}}$ is mainly controlled by the increase of the shear modulus since the transverse speed of sound $c_{\mathrm{t}} = \sqrt{G/\rho}$
is 2.4 to 2.6 times smaller than the longitudinal speed of sound $c_{\mathrm{l}}$. The overall relative variations of mechanical moduli and
Debye frequency are, however, relatively mild given the broad range of glass stabilities covered in Fig.~\ref{macro}.

\begin{figure}
\centering
\includegraphics[width=0.48\textwidth]{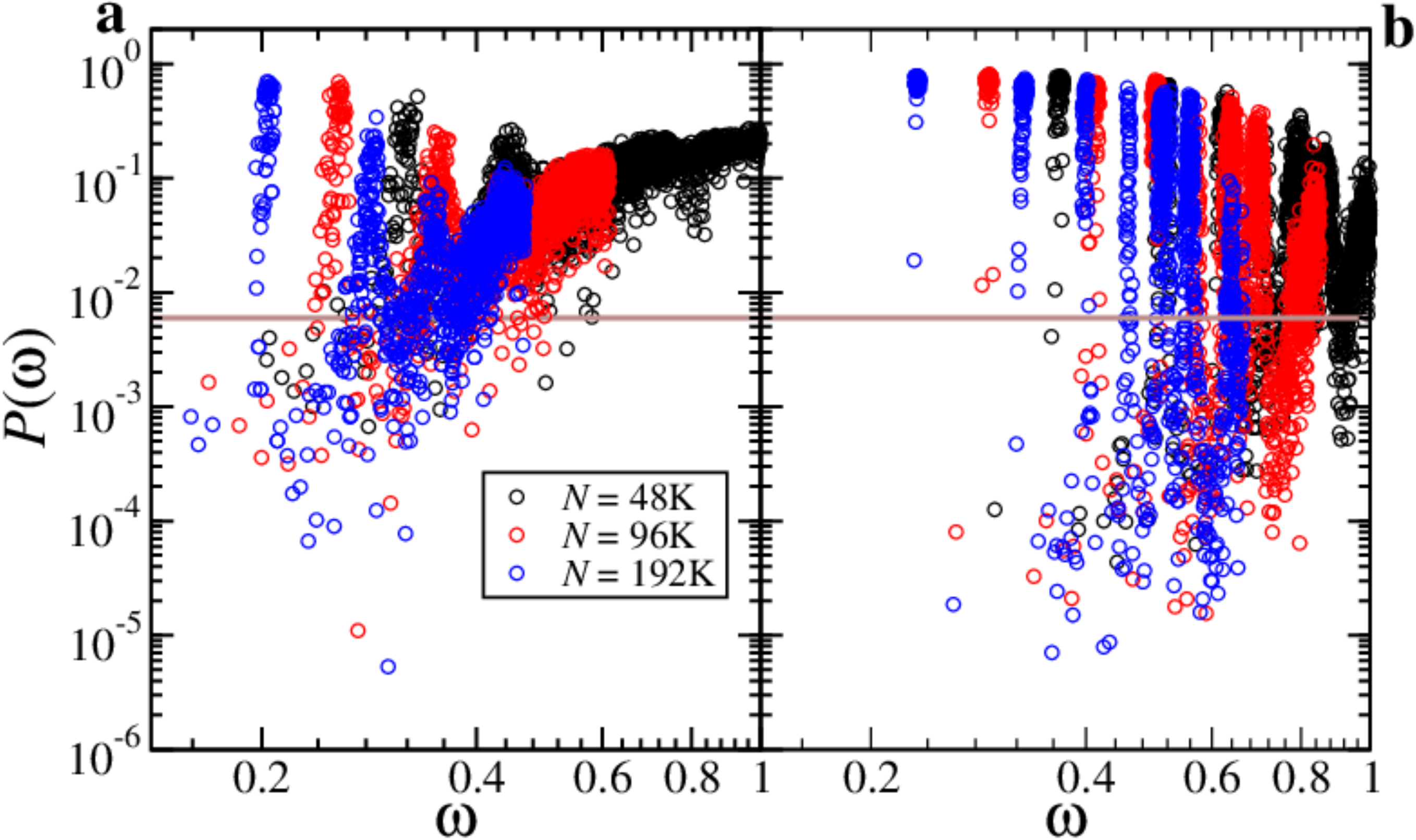}
\caption{\label{fig:fig1} \textbf{Classification of modes}. Participation ratio $P(\omega)$ as a function of frequencies $\omega$ combined from systems with $N = 48000$, $96000$ and $192000$ at parent temperatures $T_{\mathrm{p}}=0.200$ in \textbf{a} and $T_{\mathrm{p}}=0.062$ in \textbf{b}. The horizontal line marks the threshold $P_0=0.006$ between extended and quasi-localized modes.}
\end{figure}

\textbf{Classification of  quasi-localized and extended modes.} By examining the participation ratio $P(\omega)$ as a function of $\omega$
at different parent temperatures, see Fig.~\ref{fig:fig1}, we
observe all the features that characterize the $T_{\mathrm{p}}$-dependence of the
density of states. A value of $P(\omega) = 1$ indicates a mode where all the particles participate equally,
a value of $P(\omega) = N^{-1}$ indicates a mode where only one particle participates,
and a value of $P(\omega) = 2/3$ indicates a plane wave.
The sharp peaks in $P(\omega)$ at low frequencies are due to the phonon modes, with the first peak corresponding
to the first allowed transverse phonon at $\omega_{\mathrm{t}} = c_{\mathrm{t}} 2 \pi/L$, $L$ being the box length. An increase in $\omega_{\mathrm{t}}$ indicates an increase in $c_{\mathrm{t}} = \sqrt{G/\rho}$. The low frequency modes can be naturally divided into quasi-localized modes (small $P$) and extended modes (large $P$) through an appropriate thresholding procedure \cite{ikeda_pnas,ikeda_pre2018}, this decomposition becoming sharper as $L$ increases and $T_{\mathrm{p}}$ decreases. The value $P_0 = 0.006$ is appropriate, as shown in Fig.~\ref{fig:fig1}, but we checked that our results are not qualitatively affected by a reasonable change of $P_0$. As $T_{\mathrm{p}}$ decreases, phonon modes shift to larger frequencies, as expected from the evolution of the mechanical moduli, whereas quasi-localized modes become increasingly localized and well-separated from the phonons. We also checked that our results hold for small system sizes where allowed phonon modes are shifted to much higher frequencies~\cite{lerner_prl2016}.

\begin{figure}
\centering
\includegraphics[bb=2in 0in 9in 8.5in,clip=true,width=0.48\textwidth]{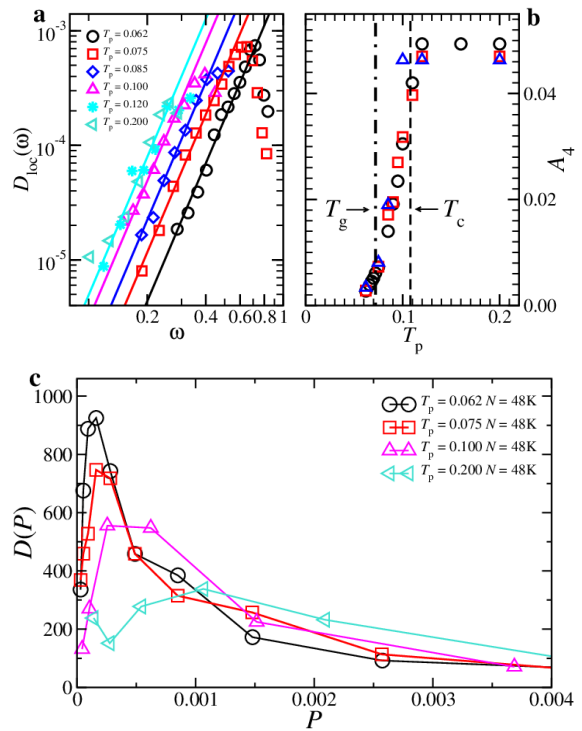}
\caption{\label{fig:fig2} \textbf{Density of states and spatial localization of quasi-localized modes}.
\textbf{a} Density of states $D_\mathrm{loc}(\omega)$ for quasi-localized modes for $N=48000$, with fits to $D_\mathrm{loc} =  A_{4} \omega^{4}$.
\textbf{b} $T_{\mathrm{p}}$ dependence of $A_4$ for $N=48000$  (circles), $96000$ (squares), and $192000$ (triangles), with the mode-coupling temperature $T_{\mathrm{c}}$ and the estimated
experimental glass temperature $T_{\mathrm{g}}$ indicated.
\textbf{c} Probability distribution of the participation ratio for quasi-localized modes in the frequency-range of the $\omega^4$ scaling for various $T_{\mathrm{p}}$.}
\end{figure}

\textbf{Properties of quasi-localized modes.} We
examined the density of states for the quasi-localized modes $D_{\mathrm{loc}}(\omega)$, which are shown in Fig.~\ref{fig:fig2}a
for a few representative $T_{\mathrm{p}}$. At low frequencies, $D_{\mathrm{loc}}(\omega) = A_4 \omega^4$ for each parent temperature
with a prefactor $A_4$ that depends on the glass stability. We show the resulting $A_4(T_\mathrm{p})$ in Fig.~\ref{fig:fig2}b.
The prefactor $A_4$ stays nearly constant for high enough $T_\mathrm{p}$, but decreases sharply when $T_\mathrm{p}$ decreases below the mode-coupling temperature $T_\mathrm{c}$.
This observation is robust against changing the system size. The decrease of $A_4$ at low $T_\mathrm{p}$ correlates well with the
evolution of shear modulus and Debye level in Figs.~\ref{macro}(c,d). We note that a study of less stable glasses \cite{Lerner2018}
found an increase in the lowest frequency of quasi-localized modes with decreasing parent temperature, which, under certain assumptions, may be related to the change of $A_4$ reported here.
A major result of our study is that the quartic law governing $D_\mathrm{loc}(\omega)$ is obeyed irrespective of the glass stability, thus extending the
validity of previous findings to the experimentally relevant regime.

In Fig.~\ref{fig:fig2}c we show the probability distribution for finding a mode with a participation ratio $P$ for the modes with $P<P_0$ for $N=48000$ particles.
With decreasing $T_\mathrm{p}$, the distribution becomes narrower and the peak position shifts to smaller $P$ values.
We find that the average participation ratio decreases with decreasing
$T_\mathrm{p}$, which is evident from Fig.~\ref{fig:fig2}c. This confirms that these modes become more localized with decreasing
parent temperature, which had been observed for less stable glasses
\cite{lerner_pre2017,lerner2018b,Lerner2018}. Since the density of states is a function of the structure of the quenched system, we conclude
that subtle local structural changes occur for $T_\mathrm{p}$ below $T_\mathrm{c}$ that strongly affect soft vibrational motion in the quenched glass.

\begin{figure}
\includegraphics[bb=2in 0in 9in 8.5in,clip=true,width=0.48\textwidth]{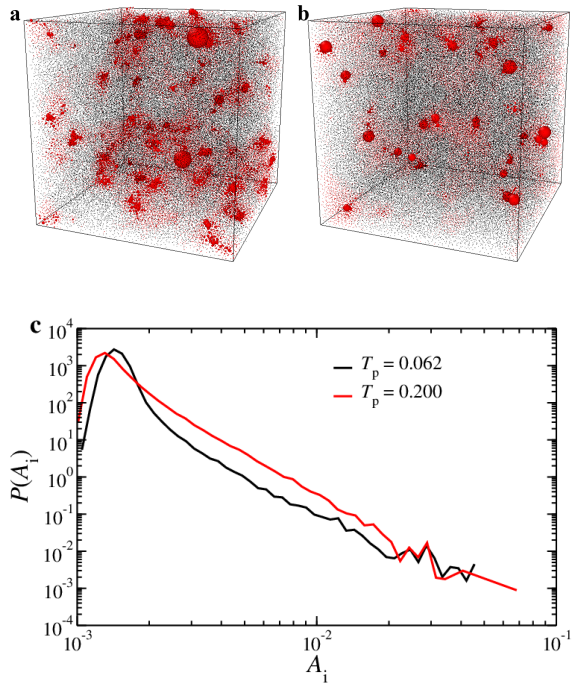}
\caption{\label{fig:snap} \textbf{Softness of quasi-localized modes}.
Snapshots obtained for $T_{\mathrm{p}}=0.200$ (\textbf{a}) and $T_{\mathrm{p}}=0.062$ (\textbf{b}) with $N=192000$. The particles are shown with their radius given by the vibrational amplitude $A(i)$ calculated from the lowest quasi-localized modes.
\textbf{c} The probability distribution of $A(i)$ for $T_{\mathrm{p}} = 0.200$ and $T_{\mathrm{p}} = 0.062$. For lower parent temperatures there is a smaller fraction of the particles with larger $A(i)$, and thus the modes are more localized.}
\end{figure}

To visualize the increasing mode localization, we define a `softness' ~\cite{rottler_prx} for particle $i$ as $A(i) = (1/M) \sum_{l=1}^M | \mathbf{e}_{l,i} |$ where the sum is taken over the $M=40$ lowest frequency quasi-localized modes for one inherent structure (we have checked that our conclusions hold when we take other values of $M=5-40$). The softness quantifies the vibrational amplitude of low-frequency quasi-localized modes. In the snapshots of Fig.~\ref{fig:snap}, particles are represented with a size proportional to $A(i)$ for (a) $T_\mathrm{p}=0.200$ and (b) $T_\mathrm{p}=0.062$. For the highest $T_\mathrm{p}$, clusters contributing to localized modes are relatively numerous, quite extended, and strongly coupled to their environment.  At the lowest $T_\mathrm{p}$, each cluster is localized around just a few particles, there are much fewer clusters, and they offer a stronger contrast with the immobile background. To quantify these observations, we measured the probability distribution of $A(i)$, Fig.~\ref{fig:snap}c. These distributions show a power law tail at large $A$ values, $P(A_i) = \lambda(T_\mathrm{p}) A_i^{-\alpha}$ with $\alpha \approx 3.7$. At low $T_\mathrm{p}$ the tail is well separated from the core of the distribution at small $A$, and mobile particles with large $A$ are better defined. There is also a pronounced decay of the probability of finding large $A$ values at low $T_\mathrm{p}$ since $\lambda(0.2) / \lambda(0.062) \approx 4.3$, which indicates a greater than four fold decrease in the number of soft particles with large vibrational amplitudes. The interpretation of quasi-localized modes as relevant glassy defects controlling mechanical and thermal properties of glasses is therefore more convincing for stable glasses than it is for conventional computer glasses.

\begin{figure}
\centering
\includegraphics[bb=2in 0in 9in 8.5in,clip=true,width=0.48\textwidth]{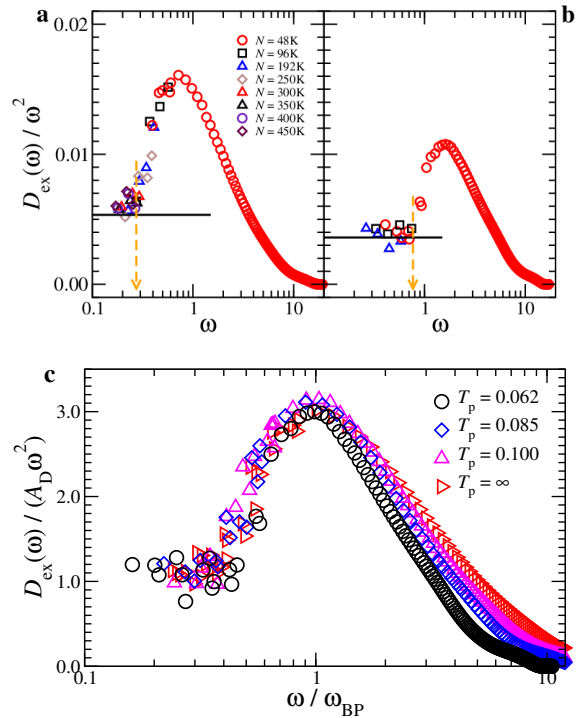}
\caption{\label{fig:fig3}  \textbf{Density of states for extended
modes.}
Reduced density of states for extended modes,
$D_\text{ex}(\omega)/{\omega}^2$, at parent temperatures $T_\mathrm{p}=\infty$ (\textbf{a}) and $0.062$ (\textbf{b}), for  systems with different sizes. The black line in each panel indicates the Debye level $A_{D}$ while the vertical arrow marks the frequency where $D_\text{ex}(\omega)/{\omega}^2$ starts to deviate from $A_\mathrm{D}$.
\textbf{c} Rescaled version of the same data using the Debye level (vertical axis) and the boson peak frequency $\omega_{\mathrm{BP}}$ (horizontal axis).}
\end{figure}

\textbf{Properties of extended modes.} Next, we examine the density of states of extended modes, $D_{\mathrm{ex}}(\omega)$, with a participation ratio greater than $P_0$.
In Figs.~\ref{fig:fig3}(a,b) we show the reduced density of states $D_{\mathrm{ex}}(\omega)/\omega^2$ for two parent temperatures. For each temperature, the Debye level is reached at low enough $\omega$ and a boson peak is observed at larger frequencies. Using our localization criterion, we find that modes near the boson peak are not localized, but this does not imply that they have a phononic character. The boson peak narrows slightly with decreasing $T_\mathrm{p}$.  The Debye level, the boson peak location, height and width all change modestly as $T_\mathrm{p}$ is varied over the entire range studied.  The changes observed in our study agree qualitatively with those found by Grigera \textit{et al.} \cite{parisi_nature}.

In Fig.~\ref{fig:fig3}c we examine scaling properties of the density of states of extended modes. We rescale $\omega$ by the boson peak frequency, $\omega_{\mathrm{BP}}$, and plot the rescaled density of states $D_\text{ex} / (A_\mathrm{D} \omega^2)$. We observe an excellent collapse on the low-frequency side of the boson peak. This shows that in this frequency range the reduced frequency dependence has a universal shape, as reported before~\cite{tanaka_nm2008}. Second, the collapse also shows that the height of the boson peak correlates with the Debye level $A_\mathrm{D}$. These results agree with experiments on molecular glass formers~\cite{Monaco2006,Baldi2009,Monaco2006a}. However,
some of the same experiments report that the boson peak position scales as the Debye frequency~\cite{Monaco2006,Monaco2006a}, which is not consistent with our results. We also find that a scaling of $\omega_{\mathrm{BP}}$ with the bulk modulus suggested in Ref.~\cite{Ruffle2010} is inconsistent with our results. Note that we study the evolution of the boson peak as a function of the preparation temperature, while experiments sometimes examine the temperature evolution of the boson peak for a given glass preparation. We also note
that a correlation between the boson peak and quasi-localized modes has been proposed by studying systems at different pressures
around the unjamming transition \cite{Shimada2018}.

\begin{figure}
\includegraphics[width=0.48\textwidth]{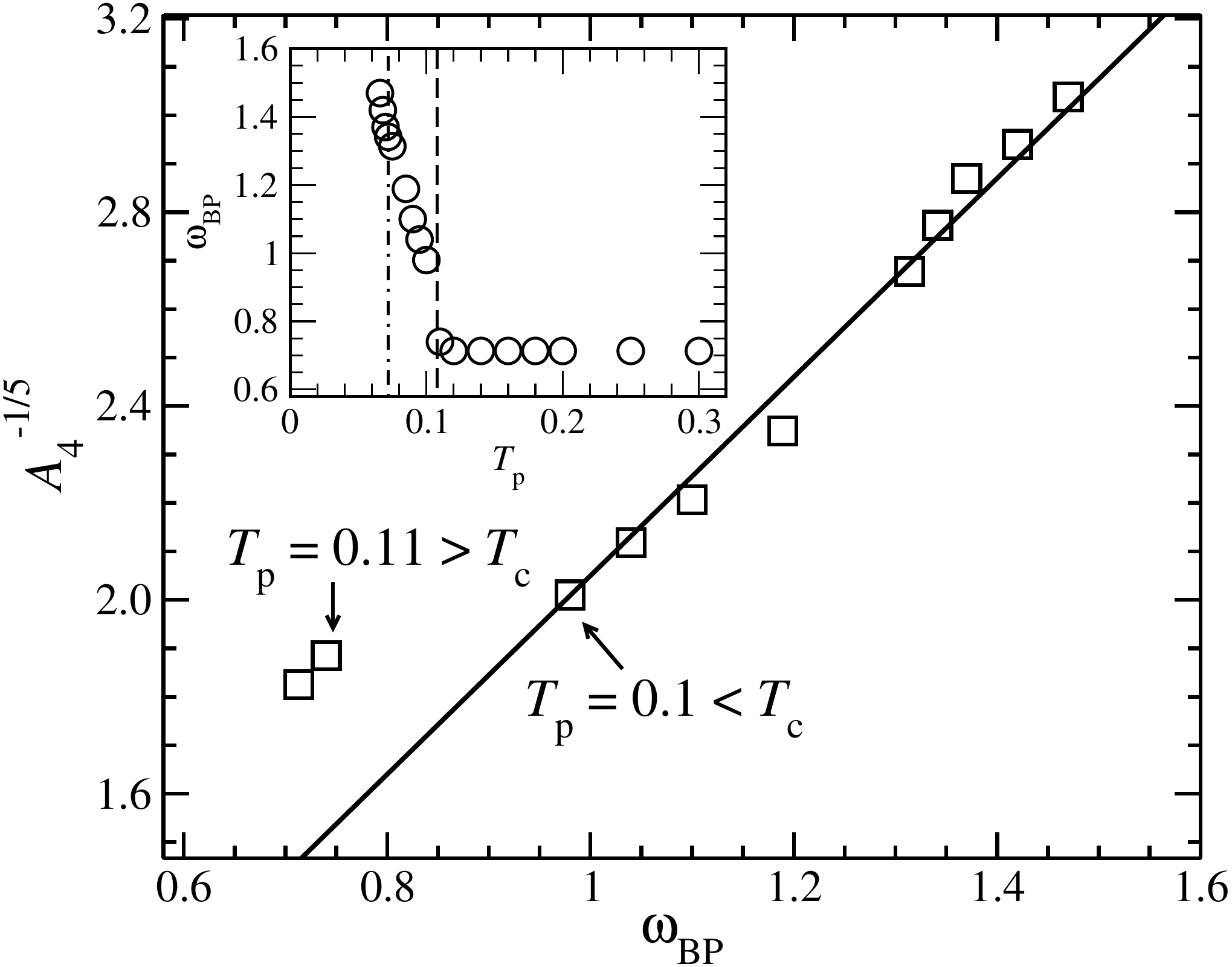}
\caption{\label{fig:fig6}\textbf{Scaling of the boson peak frequency.} The characteristic frequency $A_4^{-1/5}$ versus
the boson peak position $\omega_\mathrm{BP}$. The line is a fit $A_4^{-1/5}\sim \omega_\mathrm{BP}$
for glasses whose $T_\mathrm{p} < T_\mathrm{c}$,
which is the parent temperature range where we see an increase in $\omega_\mathrm{BP}$ with decreasing
$T_\mathrm{p}$, see inset where the vertical dashed-dotted, dashed lines mark the positions of $T_{\mathrm{g}}$ and $T_{\mathrm{c}}$, respectively.}
\end{figure}

Since the boson peak occurs in a different frequency range than the $\omega^4$ scaling of $D_\mathrm{loc}(\omega)$, it is not
clear that there could be a relationship between the boson peak and the low-frequency quasi-localized modes.  Simulations close
to jamming suggest that $A_4 \sim \omega_\mathrm{BP}^4$ \cite{ikeda_pnas}, but we do not find that this relation holds
with changing $T_\mathrm{p}$.
An alternative possibility can be obtained from dimensional analysis,
where a characteristic  frequency for quasi-localized modes can be defined as ${A_4}^{-1/5}$  \cite{Lerner2018}. We find that ${A_4}^{-1/5} \sim \omega_\mathrm{BP}$ for glasses with $T_\mathrm{p} < T_\mathrm{c}$, Fig.~\ref{fig:fig6},
but this relation does not hold for glasses created with $T_\mathrm{p} > T_\mathrm{c}$. We note that $\omega_\mathrm{BP}$ is
constant for $T_\mathrm{p} > T_\mathrm{c}$, see the inset to Fig.~\ref{fig:fig6}, and
only changes for $T_\mathrm{p} < T_\mathrm{c}$. Again we find that $T_\mathrm{c}$ marks a change in the
behavior of $D(\omega)$.
Given the relatively small changes in both $\omega_\mathrm{BP}$ and ${A_4}^{-1/5}$ over our entire range of parent temperatures studied, it is not clear that
a power law is the proper relationship between these quantities and further work is needed to verify it.

\section*{Discussion}

In summary, we report the first characterization of the vibrational density of states of computer glasses prepared over a range of glass stability that bridges the gap between ordinary simulations and experimental studies. At low frequency extended and quasi-localized modes coexist, and both types of modes evolve differently when the glass stability is varied. We find a relatively mild temperature dependence of extended modes, with a strong correlation between the Debye level and the boson peak. By contrast, quasi-localized modes evolve more strongly when $T_\mathrm{p}$ decreases below the mode-coupling temperature, but their density of states is always described by $D_\text{loc} \sim A_4 \omega^4$.
Unexpectedly, the temperature dependence of the prefactor $A_4(T_\mathrm{p})$ is more interesting than the value of the quartic exponent, which is insensitive to the degree of annealing.

The increasing localization of the modes implies that subtle yet significant changes occur in the local structure of the glass that are not reflected in the pair correlation function, which is nearly identical for parent temperatures below $T_\mathrm{c}$. Since soft modes have been linked to irreversible relaxation \cite{Widmer-Cooper_np} and rearrangements under shear \cite{rottler_prx,chen_prl,manning_prl,Zylberg_PNAS2017}, it follows that the reduction of these soft modes can have significant implications for glassy dynamics. In turn this reduction indicates that there are fewer soft spots, which should increase the strength of the glass. This hypothesis is supported by the observation that the decrease in $D_{\mathrm{loc}}(\omega)$ mirrors the increase of the shear modulus, and also correlates very well with the evolution of the ductility of the produced glasses~\cite{misaki,Ketkaew2018NC}.
Since we can now equilibrate amorphous systems at temperatures low enough so that they do not flow, another perspective would be to analyze the density of states at finite temperatures through the Fourier transform of the velocity autocorrelation function~\cite{Velocity}, or by diagonalizing the covariance matrix of displacements~\cite{Covariance}. Future studies should examine the difference between these procedures to provide insights into thermal anharmonicities of stable glasses, and more generally into their low-temperature transport properties.

\section*{Methods}
\textbf{Simulations.}
We simulate a polydisperse model glass former of
sizes between $N=48000$ and
$450000$ particles with equal mass at a number density $\rho=1.0$ \cite{berthier_prx2017}.
The interaction between two particles ${i}$ and ${j}$  is given by
$V(r_{ij}) = \left(\frac{\sigma_{ij}}{r_{ij}}\right)^{12} + v(r_{ij}) $
when their separation $r_{ij} \leq r_{ij}^c=1.25\sigma_{ij}$ and zero otherwise.
We use $v(r_{ij})=c_{0}+c_{2}\left(\frac{r_{ij}}{\sigma_{ij}}\right)^{2}+ c_{4}\left(\frac{r_{ij}}{\sigma_{ij}}\right)^{4}$,
where the coefficients   $c_{0}$, $c_{2}$ and $c_{4}$  ensure the continuity of $V(r_{ij})$ up to the second derivative at the cutoff $r_{ij}^c$.
The probability of particle diameters  $\sigma$  is $P(\sigma)=A/\sigma^3$, where $\sigma\in [0.73,1.63]$ and we use a non-additive mixing rule, $\sigma_{ij}=\frac{\sigma_{i}+\sigma_{j}}{2}(1-0.2 |\sigma_{i}-\sigma_{j}|)$. For $N\le 192000$ we use the swap Monte Carlo algorithm to prepare independent equilibrated configurations at parent temperatures $T_p$ ranging from above the onset temperature of slow dynamics ($T_\mathrm{o} \approx 0.200$) down to $T_\mathrm{p} = 0.062$, which is about $60\%$ of the mode-coupling temperature ($T_\mathrm{c} \approx 0.108$), and is lower than the estimated experimental glass temperature ($T_\mathrm{g} \approx 0.072$) \cite{berthier_prx2017}.
In addition, we also use a very high parent temperature, which we refer to as $T_\mathrm{p}=\infty$. Due to very long equilibration times for systems of $N>192000$ particles we only study systems with $N>192000$ for $T_\mathrm{p}=\infty$.

\textbf{Density of states calculation.}
Following equilibration at a temperature $T_p$, zero-temperature glasses are produced by instantaneously quenching equilibrium configurations to their inherent structures using the Fast Inertia Relaxation Engine algorithm~\cite{fire}. We then calculate the modes by diagonalizing
the Hessian matrix using Intel Math Kernel Library (https://software.intel.com/en-us/mkl/) and ARPACK
(http://www.caam.rice.edu/software/ARPACK/). We calculate all the normal modes for the $48000$ particle systems,  but only the low-frequency part of the spectrum in systems with $N > 48000$. We characterize the modes through the density of states  $D(\omega)=\frac{1}{3N-3} \sum_{l=1}^{3N-3} \delta(\omega-\omega_l)$ and the participation ratio
$P(\omega_{l})=\frac{\left(\sum_{i=1}^N |\mathbf{e}_{l,i}|^2\right)^2}{N\sum_{i=1}^N |\mathbf{e}_{l,i}|^4}$, where $\mathbf{e}_{l,i}$  is the polarization vector of particle $i$ in mode $l$ with  frequency $\omega_{l}$. For a mode localized to one particle $P(\omega) = N^{-1}$, and for an ideal plane wave $P(\omega) = 2/3$. The phonon modes occur at discrete frequencies, and care has to be taken in the binning procedure to calculate the density of states of extended modes, $D_{\mathrm{ex}}(\omega)$. To perform this calculation, we determine the phonon frequencies from the peak positions of the participation ratio versus frequency, and tune the bin size to smooth $D_{\mathrm{ex}}(\omega)$.To obtain the shear modulus $G$ and the bulk modulus $B$ we use the method described in Ref.~\cite{OHern2003}.
\\

\section*{Data Availability}
All data will be available from the authors upon request.

\section*{Acknowledgements}

We thank J.-L. Barrat for discussions.
L.W., E. F., and G.S. acknowledge funding from NSF DMR-1608086.
This work was also supported by a grant from the Simons foundation (No. 454933 L. B.). L. W. and P. G. acknowledge support from the National Natural Science Foundation of China (No. 51571011), the MOST 973 Program (No. 2015CB856800), and the NSAF joint program (No. U1530401). We acknowledge the computational support from Beijing Computational Science Research Center.
\\
\section*{Author contributions}

L. B., G. S., and E. F. designed the project. L. W., and A. N. performed  numerical simulations. L. W.,  A. N., L. B., P. G.,  G. S., and E. F.  contributed to the analysis of data. L. W., L. B., G. S., and E. F. wrote the paper.

\section*{Additional information}
Correspondence and requests for materials should be addressed to E. F. (flennere@gmail.com) or P. G. (pguan@csrc.ac.cn).

\textbf{Competing interests}: The authors declare no competing financial or non-financial interests.

\end{document}